# Spontaneous Origin of Topological Complexity in Self-Organizing Neural Networks

George Chapline

Lawrence Livermore National Laboratory
Livermore, CA 94550
chapline1@llnl.gov

## Abstract

Attention is drawn to the possibility that self-organizing biological neural networks could spontaneously acquire the capability to carry out sophisticated computations. In particular it is shown that the effective action governing the formation of synaptic connections in models of networks of feature detectors that encorporate Kohonen-like self-organization can spontaneously lead to structures that are topologically nontrivial in both a 2-dimensional and 4-dimensional sense. It is suggested that the appearance of biological neural structures with a nontrivial 4-dimensional topology is the fundamental organizational principle underlying the emergence of advanced cognitive capabilities.

# 1. Introduction

Although the underlying organizational principles of the brain that provide humans with their unique intellectual capabilities are as yet unknown, it is reasonable to surmise that the ability of the human brain to process both sensory data and stored information in a sophisticated manner is intimately related to the topological complexity of the synaptic connections in the cerebral cortex. Just how the patterns of synaptic connections in the cerebral cortex are established is somewhat mysterious, particularly in view of the fact that the human genome doesn't contain nearly enough genetic information to specify how all the neurons in the cortex should be interconnected. It is also noteworthy that humans possess mental capabilities such as reading that could not have evolved by natural selection. In this paper we would like to suggest that the topological structures in the cerebral cortex that endow humans with their superior intelligence might have arisen spontaneously as a result of random excitations of the developing brain.

Although it is clear that the main function of an animal's brain is to processes information, it is also clear that biological brains are radically different from conventional computers [1]. The most obvious reason for this difference is the fact that biological brains employ much larger networks of neurons to carry out computations than the networks used in even the largest man-made parallel computers. However, a more subtle difference has to do with the way the endpoint of computations is specified. Conventional computers make use of a "program" to organize computations, while biological brains to a large extent self-learn how to correlate desirable outputs with sensory and/or memory inputs. In both conventional computers and biological brains the endpoint of a computation is a quiescent or stationary state. of the system. However, in contrast with conventional computers biological brains in general do not have the capability of instantly reprogramming themselves to change the output state that corresponds to a given input state. Therefore biological brains might be characterized as physical systems whose stationary states are in one to one correspondence with pairs of inputs and outputs. In the following we would like to introduce the notion that the requirement that specified inputs should be correlated certain outputs can be interpreted to mean that only the boundary conditions for the physical system have dynamical significance. In other words, the only observable quantities for the system are purely topological [2]. The global behavior of such a system will be determined not just by the boundary conditions but also by the topology of the space in which the system lives. In four dimensions the boundary conditions are the inputs and outputs, and one of our basic ideas will be that the topology of a certain four dimensional space might play much the same role in the cerebral cortex as the program in an ordinary computer in determining how outputs should be correlated with inputs.

A physical system with purely topological excitations can be described either in terms of topological invariants or stochastic equations [3]. These stochastic equations are suggestively similar to the time evolution equations for the network of resistors and operational amplifiers introduced by Hopfield [4] as a model for a biological neural network. These two types of systems have the common feature that the only stationary states are ground states; although in the case of a Hopfield network the various ground states do not have an obvious topological interpretation. On the other hand it is clear that the topology of a nonplanar graph whose edges correspond to synaptic connections plays a crucial role in the functioning of the Hopfield network [4]. Actually one of our objectives in this paper is to develop an approximation perhaps more suitable for describing the cerebral cortex of animals that allows sensory data to be stored in the form of three dimensional manifolds. Psychological experiments show that three dimensional smooth manifolds play a central role in the human brain in providing internal representations of the external world [5] In this case the underlying space whose topology is important for determining behavior will be a four dimensional manifold.

Our main interest in this paper will be to show how non-trivial topological structures for the neurons and synapses in biological brains can arise. We will follow an approach initiated by Little [6] and Hopfield [7]. Namely, one replaces the biological system of neurons and synapses with a previously studied statistical mechanical model, which is then analyzed using the standard techniques of mathematical physics. The statistical mechanical model for neurons and synapses originally introduced by Hopfield [7] was based on the theory of spin glasses. Hopfield mainly used numerical techniques to analyze the behavior of a spin glass as a model for neurons and synapses. However, later work by Gardner [8] demonstrated the power of purely analytical techniques to analyze spin glass models for neural networks. In particular Gardner used path integral techniques to determine how many patterns could be stored by varying the connection strengths in a spin glass with non-local interactions. Our work might be viewed as an extension of Gardner's work in which we study the effect of topologically different connectivities in a network by replacing the spin glass model with certain quantum gravity models. We do not mean to imply by the introduction of these models that we believe quantum mechanics plays any significant role in biological brains. To the contrary biological brains are almost certainly essentially classical systems. However we would like to introduce the formal structure of Euclidean quantum gravity as a useful tool for describing the spontaneous appearance of various topologies for the synaptic connections. The role played by "quantum fluctuations" in these Euclidean quantum gravity models is just to simulate the effect on the development of synaptic connections in the cerebral cortex of random signals generated in the mid brain before birth or in the environment after birth. Another advantage of using a quantum gravity model for a biological brain is that this enables us to make contact with well known examples of topological field theories.

For the purposes of this paper we shall be concerned not so much with what particular statistical mechanics models might be most accurate for describing the detailed behavior of the system of neurons and synaptic connections in biological brains, as finding models which show how nontrivial topological structures in the cerebral cortex might in principle have arisen. In the following sections we will introduce three analytically solvable statistical mechanics models of increasing sophistication which illustrate how nontrivial topological structures in a self-organizing neural network might arise. The first model describes how feature detectors within a single layer of the cerebral cortex can self-organize to produce vortex structures similar to the pinwheel-like singularities in the organization of orientation preference columns that have been observed in the primary visual cortex of monkeys [9]. In section 3 we will interpret these pinwheel-like singularities as evidence that the orientation sensitive neurons within a single layer of the visual cortex are interconnected in a topologically nontrivial way, and introduce a two dimensional quantum gravity model which may be a more appropriate model for a single layer of feature detectors in the cerebral cortex. In section 4 a four dimensional quantum gravity model is introduced which provides a description for a foliation of feature detecting networks. This last model illustrates, at least in principle, how local self-organizing dynamics can spontaneously give rise to a cooperative assembly of neural circuits, each of which is specialized to detect different features, but collectively can combine the information from different kinds of feature detectors to make decisions.

## 2. Self-organization of Orientation Columns

Self-organization of synaptic connections so that neurons that process sensory data from physically adjacent parts of the environment are close together in the cortex may be a common phenomenon in brain development [10]. For example, numerical simulations based on Kohonen's self-organizing map algorithm have successfully reproduced qualitative features of the organization of orientation preference and ocular dominance columns within each hypercolumn of the visual cortex of the macaque monkey [11]. A particularly interesting result of these simulations is the occurrence of vortex-like singularities in the arrangement of orientation columns. The authors of ref.11 interpreted the occurrence of these vortex singularities in terms of singularities that occur in dimension reducing maps. In this and the next section we would like to offer two alternative topological interpretations for these singularities, based on a simple neural network model for self-organization of orientation preference columns.

Our basic network consists of N feature detectors such that each feature detector is connected to three neighboring feature detectors. The assumption of three connections per neuron is made for convenience since models where the feature detectors are allowed to connect to larger numbers of neighbors lead to similar results. Also in this report we will concentrate on the case where each feature detector is characterized by an angle $w$. For example, the orientation sensitive neurons within the primary visual cortex are characterized by a preferred orientation which denotes the stimulus orientation which gives the strongest response.

We wish to develop a theory for how neurons specialized to detect an orientation angle are organized within a single layer of the cerebral cortex. As our starting point we consider maps that assign to each environmental orientation $\phi$ a location $r$ within our network of N feature detectors. Following Kohonen [10] we will assume that brain development can be modeled by assuming that the maps of interest are "self-organizing". That is initially each feature detector is assigned a random orientation $w(r,0)$, and each orientation $\phi$ in the environment is mapped to that feature detector r whose orientation $w(r,0)$ is closest to $\phi$. Thereafter the orientation of the feature detector located at r evolves according to a rule of the form

$$w(r,t+1) = w(r,t) + h(r-s)[\phi - w(r,t)], \qquad (1)$$

where h(x) is typically assumed to be a Gaussian function peaked at x=0. In the following the function h(r-s) will be replaced by the rule that each feature detector is connected only to its three nearest neighbors. The location s in (1) corresponds to the feature detector whose orientation $w(s)$ is closest to $\phi$. Thus the developmental process is modeled as a Markov process whose states are the sets $\{w(r)\}$ of possible states of the feature detectors, and where the transition probabilities are determined by probabilities of occurrence in the environment of various orientations $\phi$. In order to construct an analytical model of this developmental process it will be useful to introduce an energy functional E[$w$] that satisfies

$$< P(\phi)\delta w > = -grad_w E \qquad (2)$$

where $\delta w = w(r,t+1) - w(r,t)$ and P($\phi$) is the probability distribution for the orientations of the environmental stimuli. Neglecting certain mathematical subtleties, the required energy functional is [12]

$$E[w] = \frac{1}{2} \sum_{<r,s>} \sum_{\phi \in R} P(\phi) |\phi - w(r,t)|^2 \qquad (3)$$

where the sum over <r,s> runs over nearest neighbor connections and $R(r)$ is the receptive field of the feature detector located at r; i.e. the union of all environmental stimuli that are closer to $w(r, t)$ than any other $w(s,t)$, where $s \neq r$.

Given an energy functional that satisfies (2) there are standard techniques that one can use to describe the stochastic evolution of the organization of our neural network. However in the following our only interest in how the organization of feature detectors evolves with time will be limited to noting that under the influence of the random variable $\phi(t)$ the system relaxes to an asymptotic state characterized by a stationary probability distribution for various final configurations $\{w(r)\}$. The statistical properties of our network of feature detectors in this stationary state can be derived from a "partition function", which is a sum over all possible stationary state configurations weighted with the Boltzmann factor exp(-E[w]). If we assume that the stochastic evolution of our network is governed by an energy functional of the form (3) then this partition function has the form:

$$Z = \sum_L \kappa^F \prod_{i=1}^{F} \int_o^{2\pi} dw(r_i) e^{-\frac{K}{2} \sum_{<i,j>} |w(r_i) - w(r_j)|^2} \qquad (4)$$

where $\kappa$ and K are constants, the sum over L means a sum over triangular lattices, and the indices i and j refer to orientation sensitive neurons located at the centers of the triangles in this lattice (note that N is the number of faces of the lattice L). For large numbers of faces the triangular lattices L can be thought of as triangulations of 2-dimensional surfaces, and in the limit $N \to \infty$ the sum over triangular lattices in (4) becomes a sum over smooth 2-dimensional surfaces. In this limit the partition function (4) becomes

$$Z = \int Dw(\sigma) \exp(-S), \qquad (5)$$

where $(\sigma_1, \sigma_2)$ are the coordinates of a point on the smooth surface and the continuum action S is given by

$$S = \frac{K}{2} \int d^2 \sigma \partial_\alpha w \partial_\alpha w + \lambda. \qquad (6)$$

The constant $\lambda$ in (6) replaces the constant $\kappa$ and plays the role of energy per neuron. Partition functions using classical actions similar to (6) were originally introduced as theories of matter coupled to 2-D quantum gravity [13]. In particular, the quantum theory defined by (5) describes the coupling of 2-D quantum gravity to a single scalar field, and has been intensively studied by mathematical physicists [14]. If one assumes that this scalar field represents a periodic variable, then it turns out that there is a phase of the theory where the dynamics is essentially the dynamics of 2-D quantum gravity.

Another interpretation [14] of the partition function (5) is that it represents a relativistic string moving on a 2-dimensional surface - in mathematical terms this means holomorphic mappings from an arbitrary 2-dimensional manifold onto a fixed 2-dimensional manifold. In this string interpretation the angle variable w becomes a complex variable by the addition of a second real variable representing the local magnification of the mapping. It is worth noting that this result is consistent with the theorem [15] that for maps of 2-dimensional surfaces onto 2-dimensional surfaces the stationary state of Kohonen's algorithm is a holomorphic (or anti-holomorphic) map. Thus we arrive at the general result that the complex coordinate w is a function of either z=x+iy or z=x-iy. Of particular interest are solutions where w(z) has the form

$$w(z) = \sum_i m_i \operatorname{Im} \ln(z - Z_i), \tag{7}$$

where the $m_i$ are integers. In general one must have $\sum_i m_i = 0$. In fact these solutions are just the vortex configurations of the 2-dimensional XY model discovered by Kosterlitz and Thouless [16]. Substituting the configuration (7) into the action (6), the path integral (5) assumes a form identical to the partition function for a 2-dimensional Coulomb gas, with $1/\pi K$ playing the role of temperature:

$$Z_v = \sum_{m_i, Z_i} \exp\left[-\pi K \left\{ \sum_{i \neq j} m_i m_j \ln \frac{l}{|Z_i - Z_j|} \right\} \right] \tag{8}$$

As was first pointed out by Kosterlitz and Thouless [16] a 2-dimensional Coulomb gas has a phase transition which implies that at low temperatures the vortex-anti vortex pairs in the system (8) are bound together, while at high temperatures they are dissociated. Although the exact dependence of the string theory partition function (5) on temperature is not quite the same as for the XY model, it can be shown [17] that the basic picture of a Kosterlitz-Thouless (KT) phase transition holds in string theory. Applied to our network of orientation sensitive neurons the existence of a KT phase transition means that if the constant K is less than a certain critical value then vortex-like configurations should appear. It is of course interesting that vortex-like configurations are a prominent feature of the topographic organization of orientation preference columns in the primary visual cortex of primates.

## 3. 2-D Quantum Gravity Model

As mentioned in the last section our theory of self-organized orientation columns can also be interpreted as a theory of 2-dimensional quantum gravity coupled to a scalar field representing orientation preference. Inspection of the action (6) reveals that when the coupling constant K is very small one can neglect kinetic variations in the scalar field. This suggests an alternative approach to understanding the appearance of vortex configurations in the pattern of orientation columns. To begin with we assume that the scalar field associated with orientation preference is frozen into a final configuration and no longer need be treated as a fluctuating field. On the other hand the "quantum gravity" degrees of freedom are still active. This means that when F is large and the coupling K is less than the critical value $K_c = 2/\pi$ our theory of self organized feature detectors effectively becomes a theory of random surfaces, where points of the random surface are labeled with a fixed orientation preferences. This interpretation of the "weak coupling" phase for our network of feature detectors leads us to another interpretation for the vortex configurations of orientation columns.

For a fixed number N of feature detectors the activation of "quantum gravity" degrees of freedom essentially means that the 2-dimensional surface approximated by a triangular lattice that is dual to the network of feature detectors will change with time. This does not mean that the local connections between feature detectors will change, but the global way this network of feature detectors folds back on itself will be allowed to vary (the network can be thought of as a discretization of a 2-dimensional surface with varying topology). In fact when both N is large and the constant $\kappa$ approaches a certain critical value, the typical topological genus occurring in the sum (4) can become very large [18]. It is not hard to show that the projection of a 2-dimensional surface with many handles onto a topologically trivial two dimensional surface that one might naively associate with a single layer of the visual cortex will necessarily lead to singularities that look like the KT vortices (7). In particular one can make use of the Riemann- Hurwitz mapping theorem [19], which relates the topological indices of a holomorphic mapping between two 2-dimensional manifolds to the topological genus of each surface. The topological indices that appear are the ramification indices $n_i$ which describe the behavior of the mapping near singular points and the winding number n which describes the number of times the mapped surface is covered by the mapping. In our case we are interested in mapping the surface associated with our network of feature detectors to the disc-like region constituting a single layer of a hypercolumn within the visual cortex. In the case where one is mapping a surface of genus g to a disc the Riemann-Hurwitz relation reduces to the simple formula

$$\sum_i (n_i - 1) - n = 2g - 2 \quad . \tag{9}$$

where the sum runs over all singular points. If g is positive and large then it follows that there must be some $n_i > 0$. In addition, the sum $n_i$ must be large, so that if there are only a few singular points per hypercolumn the $n_i$ themselves must be large. Indeed it is reasonable to assume that the $n_i$ have magnitudes on the order of the number of orientations that the human brain can distinguish; i.e. a few thousand. Since near a singular point the mapping can be approximated as $z^{n_i}$ [19], these singular points begin to look a lot like the pinwheel-like patterns of orientation preference columns observed in the visual cortex of monkeys [9]. Therefore we are led to the suggestion that the unusual pattern of orientation preference columns observed in the visual cortex of monkeys may be a signal that the orientation sensitive neurons within a single layer of the visual cortex are interconnected in a topologically nontrivial way. Furthermore since the sum (4) automatically includes topologically nontrivial surfaces, a nonlocal pattern of synaptic connections will appear spontaneously as a result of self-organization.

**4. 4-D Quantum Gravity Model**

Although the cerebral cortex is layered the neurons in different layers interact with each other. More generally, one of the most characteristic features of mental processes is that they involve cooperation of neural circuits at different locations within the cerebral cortex. These different neural circuits typically are specialized for recognition of different aspects of sensory inputs. In this section we will show how the theory of self-organization of orientation preference columns developed in section 2 can be extended to describe a foliation of feature detecting networks, where each network in the foliation contains slightly different feature detectors. We begin by slightly altering the lattice version of the continuum theory of section 2.

In order to recognize the fact that the orientation preferences w[r] are periodic variables one can replace the exponential link factors in the partition function (4) with a Villain link factor $\sum_{m=-\infty}^{\infty} \exp[-\frac{K}{2}(w + 2\pi m)^2]$ , where the sum over m insures that the link factor is periodic under w-> w + 2 . Now the partition function (4) has the form

$$Z = \sum_L \kappa^F \prod_{i=1}^{F} \int_o^{2\pi} dw(r_i) \prod_{<ij>} \sum_{m_{ij}} e^{-\frac{K}{2}[w(r_i)-w(r_j)+2\pi m_{ij}]^2}, \qquad (10)$$

where as before the sum over L runs over triangular lattices and the feature detecting neurons are situated on the faces of this lattice.  In this lattice model a vortex is associated with a particular vertex of a triangular lattice and has a topological charge

$$M = \sum_{loop} m_{ij} \qquad (11)$$

where the sum runs over the links of the neural network surrounding the particular vertex of the triangular lattice. This discrete form of the string theory partition function is instructive because it shows that the topological charges of the KT vortices are formally identical with quantized magnetic fluxes. Indeed if we were to introduce real magnetic fields into the lattice theory (10) we would obtain an antiferromagnetic version of the 2-dimensional XY model, that again contains topological excitations [20]. Such a model has much in common with models of spin glasses, which it may be recalled was the inspirations for Hopfield's original neural network [7].

One may now make use of a trick previously introduced by the author [21] for generalizing a 2-dimensional theory of topological vortices to 3-dimensions. Namely, one replaces the effective magnet fields in (9) by a magnetic field with many "colors". The previous holomorphic mapping condition now becomes the condition

$$(D_x - iD_y)W = o \qquad (12)$$

where $D_\alpha = -i\partial_\alpha + [A_\alpha, A_\alpha$ is the nonabelian gauge potential, and W is an multi-component field that we wish to use to describe the feature preferences in a foliation consisting of $F$ layers of feature detectors. The choice of magnetic field strengths is somewhat arbitrary; however, one elegant way to maintain the topological character of the theory is to replace the string action (6) with the topological action [22]:

$$S_{top} = \int d^2\sigma Tr \varepsilon^{\alpha\beta} \{\tfrac{1}{4} F_{\alpha\beta}[W^+, W] - D_\alpha W^+ D_\beta W\} \qquad (13)$$

where B = $F_{12}$ and W are assumed to belong to the adjoint representation of the Lie group SU($F$). The classical equations of motion corresponding to the action (13) can be solved exactly [23]. Moreover in the limit $F \to \infty$ where the number of different kinds of features becomes very large the solution to these equations describes the geometry of a certain kind of 4-dimensional manifold [21]. Remarkably the geometry and topology of this manifold can be expressed in terms of the magnetic potential generated by magnetic monopole-like topological excitations. Indeed in the limit $F \to \infty$ the effective magnetic field will be given by

$$B(X) = \sum_k grad_{X_k} \left\{ \frac{l}{|X - X_k|} \right\} \tag{14}$$

where the sum runs over the positions of the monopoles and the third coordinate specifies the layer of the foliation where the monopole is located. It can be shown [24] that the magnetic monopole-like objects in (14) endow the 4-dimensional manifold with certain nontrivial topological characteristics, and the existence of a smooth metric for the 4-dimensional manifold requires the introduction of a periodic time variable.

In the present context of trying to find a neural network model for the cerebral cortex the "magnetic field" introduced in eq.(14) is merely a formal device for relating the feature detectors and synaptic connection strengths in different layers of a foliation of feature detecting networks. According to this model the nature of the feature detectors varies smoothly from layer to layer and is self-organized according to (12) within each layer. One way of interpreting the significance of the solution (14) is to make use of the remarkable fact that this solution can be related [24] to the geometry of a 4-manifold. This means that overall the feature detectors in our model are organized so that the features generate a smooth 3-dimensional manifold. While this may appear at first sight to be a very artificial form of self-organization, it is quite plausible that the evolutionary importance of forming 3-dimensional visual images might lead biological systems to habitually organize features to form 3-dimensional manifolds. Whether these 3-dimensional manifolds correspond to the solution (14) is an open question; however, it is interesting to note in this connection that it was proposed some time ago [25] that 3-D visual images are constructed from a series of 2-D images. Moreover as noted earlier psychological experiments [5] suggest that in the human brain organization of sensory data into 3-dimensional manifolds occurs quite generally.

In a manner somewhat analogous to the 2-dimensional case topological complexity appears spontaneously in our 4-dimensional model because under certain conditions the monopole gas corresponding to (14) consists of dissociated monopole- antimonopole pairs. This implies that the feature detectors in our foliation of feature detecting networks are interconnected in a topologically nontrivial and subtle way. Furthermore it is especially interesting to note that because the topology of our foliation of networks is nontrivial in a 4-dimensional sense, "time" must play an essential role in establishing the cooperative behavior of this system. Actually the fourth dimension for our topologically non-trivial 4-manifold is not ordinary time (cf. ref. 24), but instead is a periodic variable that is used to tie together the feature representations of different 2-dimensional layers to form a smooth manifold. The natural appearance of a periodic time variable in our model for a foliation of networks of feature detecting neurons is certainly intriguing in view of the suggestion [26] that visual awareness and other aspects of consciousness are the result of the rhythmic and synchronized firing of neurons in the different cortical areas concerned with the recognition of a particular object. Indeed one is very tempted to identify the periodic time coordinate of the topological 4-manifold associated with the effective magnetic field (14) with the 40-Hertz rhythm in the brain that is widely believed to be involved with consciousness.

## 5. Conclusion

One of the central mysteries of neural science is how the neurons and synapses in the human brain become organized to perform sophisticated cognitive functions. This mystery is deepened by the fact that the human genome fails by many orders of magnitude to contain enough information to specify how the immense number of neurons in the human brain should be interconnected. Evidently some principle of self-organization must be at work. Kohonen's self-organizing maps have a certain physiological plausibility [27], but the question of why Kohonen-like self-organization leads to sophisticated mental capabilities has remained unanswered. In this paper we have shown that Kohonen-like self-organization can in fact spontaneously lead to topological nontrivial structures of synaptic connections similar to those in a Hopfield network. Although there is at the present time no direct evidence that the cerebral cortex makes use of Hopfield networks, there is circumstantial evidence that Hopfield-like recurrent networks play important roles in many cognitive skills such as word recognition [28].

Our general idea that cognitive processes are associated with topologically non-trivial structures of the cerebral cortex is consistent with the well known fact that mental activities typically involve a number of different regions of the cortex. That is mental activities are inherently non-local; which is of course a notable feature of Hopfield-like recurrent networks. Our 2-D quantum gravity model for a biological neural network combines the topology of a Hopfield network with the self-organizing property of a Kohonen maps. This model suggests that a good way to solve one dimensional data fusion problems would be to twist a Kohonen-like network so that it resembles a Hopfield network. Similarly our 4-D quantum gravity model suggests that a good way to fuse 2-dimensional data representations would be to twist a multilayer self-organizing network, and in addition introduce a periodic internal time variable which serves the purpose of tying together the data representations of the neural circuits in each layer of the network so that together the different circuits are representing the same object in a seamless way. This way of combining sensory data, which might be called *topological data fusion* , provides a new paradigm for using neural networks to solve all types of data fusion problems.

The fact that our 4-D quantum gravity model ipso facto corresponds to "seamless" data fusion also suggests why biological brains may have evolved so as to contain neural structures that are topologically nontrivial in a 4-dimensional sense. In other words the appearance of neural structures which are topologically non-trivial in a 4-dimensional sense may well be the fundamental organizational development which led to the more advanced cognitive capabilities associated the cerebral cortex of vertebrates. In addition 4-dimensional topological data fusion may be from a mathematical point of view the organizational development underlying conscious awareness. Indeed the suggestive similarity between the periodic time in our 4-D quantum gravity model and the 40 Hertz oscillations generated in the thalamus leads us to agree with the suggestion of Crick and Koch that this quasi-periodic rhythm is the key to understanding conscious awareness.

A final and exceedingly important inference to be drawn from our quantum gravity models is that the detailed pattern of synaptic connections in the cerebral cortex is not predetermined. It is possible that the general level of topological complexity is genetically predetermined as a result of genome specification of those physiological characteristics that affect the levels of random excitation of the developing brain, or alternatively self-organization parameters corresponding to the parameters $k$ and $\lambda$ that occur in our 2-D quantum gravity model. However once an initial level of topological complexity has been achieved the brain fine tunes the synaptic connections by self-learning in order to perform desired neural computations. As a "practical" application of these ideas it might be noted that our simple mathematical models suggest that biological brains can acquire advanced cognitive capabilities only if the developing brain is exposed to a certain level of stimulation. Indeed the models described in this paper strongly suggest that environmental stimuli during infancy play an important role in the development of sophisticated mental capabilities.